\documentclass[journal]{IEEEtran}
\ifCLASSINFOpdf
\else
\fi

\usepackage{graphicx}
\usepackage{bm}
\usepackage{amsfonts}
\usepackage{amsmath}
\usepackage{amssymb}
\usepackage{times}
\usepackage{subfigure}
\usepackage{latexsym,bm,amsmath,amssymb} 
\usepackage{CJK}
\usepackage{hhline}

\usepackage{multirow} 
\usepackage{amsmath}
\usepackage{xcolor}
\usepackage{epstopdf}
\usepackage{cite}
\usepackage[noend]{algpseudocode}
\usepackage{algorithmicx,algorithm}

\usepackage[nobiblatex]{xurl}
\begin{document}
%
\title{Faster-than-Nyquist Signaling for Next-Generation Wireless: Principles, Applications, and Challenges}

\author{Shuangyang Li,~\IEEEmembership{Member,~IEEE,}
Melda Yuksel,~\IEEEmembership{Senior Member,~IEEE,} 
Tongyang Xu,~\IEEEmembership{Member,~IEEE,} 
Shinya Sugiura,~\IEEEmembership{Senior Member,~IEEE,}
Jinhong Yuan,~\IEEEmembership{Fellow,~IEEE,}
Giuseppe Caire,~\IEEEmembership{Fellow,~IEEE,}
and Lajos Hanzo,~\IEEEmembership{Fellow,~IEEE}
\\
\emph{(Invited Paper)}

\thanks{The work of S. Li was supported in part by the European Union’s Horizon 2020 Research and Innovation Program under MSCA Grant No. 101105732 - DDComRad. The work of M. Yuksel was supported in part by the Scientific and Technological Research Council of Turkey, TUBITAK, under grant 122E248.
The work of S. Sugiura was supported in part by Japan Science and Technology Agency (JST) ASPIRE Program (Grant JPMJAP2345), and in part by JST FOREST (Grant JPMJFR2127). The work of G. Caire was supported by BMBF Germany in the program of ``Souverän. Digital. Vernetzt.'' Joint Project 6G-RIC (Project IDs 16KISK030).
The financial support of the following Engineering and Physical Sciences Research Council (EPSRC) projects is gratefully acknowledged: Platform for Driving Ultimate Connectivity (TITAN) under Grant EP/Y037243/1 and EP/X04047X/1; Robust and Reliable Quantum Computing (RoaRQ, EP/W032635/1); Integrated Waveform and Intelligence (IWAI) under grant EP/Y000315/2; India-UK Intelligent Spectrum Innovation ICON UKRI-1859; PerCom (EP/X012301/1); EP/X01228X/1; EP/Y037243/1; } 
\thanks{
S. Li and G. Caire are with the Faculty of Electrical Engineering and Computer Science, Technical University of Berlin, Berlin 10587, Germany (e-mail:\{shuangyang.li, caire\}@tu-berlin.de).\\
M. Yuksel is with the Department of Electrical and Electronics Engineering, Middle East Technical University, Ankara 06800, Turkey (e-mail: ymelda@metu.edu.tr).\\
T. Xu is with the Department of Electronic \& Electrical Engineering, University College London, WC1E 7JE London, U.K. (e-mail: tongyang.xu.11@ucl.ac.uk).\\
S. Sugiura is with the Institute of Industrial Science, The
University of Tokyo, Tokyo 153-8505, Japan (e-mail: sugiura@iis.u-tokyo.ac.jp).\\
J. Yuan is with the University of New South Wales, Sydney, 2052 NSW, Australia (e-mail: j.yuan@unsw.edu.au).\\
L. Hanzo is with the School of Electronics and Computer
Science, University of Southampton, SO17 1BJ Southampton, U.K. (email: lh@ecs.soton.ac.uk).
}
}

\maketitle

\begin{abstract}
Future wireless networks are expected to deliver ultra-high throughput for supporting emerging applications. In such scenarios, conventional Nyquist signaling may falter. As a remedy, faster-than-Nyquist (FTN) signaling
facilitates the transmission of more symbols than Nyquist signaling without expanding the time-frequency resources. 
We provide an accessible and structured introduction to FTN signaling, covering its core principles, theoretical foundations, unique advantages, open facets, and its road map. 
Specifically, we present promising coded FTN results and  highlight its compelling advantages in integrated sensing and communications (ISAC), an increasingly critical function in future networks. We conclude with a discussion of open research challenges and promising directions.
\end{abstract}


%
\IEEEpeerreviewmaketitle

\section{Introduction}
As wireless networks evolve toward 6G and beyond, the thirst for increased throughput continue to strain the spectral resources of existing systems. While emerging technologies such as millimeter-wave (mmWave) and terahertz (THz) communications promise expanded bandwidth, the reality remains that spectrum is fundamentally finite and expensive. This pressure calls for a reevaluation of physical layer signaling schemes to push the limits of spectral efficiency without relying solely on wider bandwidths.

Conventional digital communication systems have traditionally adhered to the Nyquist criterion, which ensures intersymbol interference (ISI)-free transmission in the absence of channel-induced dispersion. This Nyquist signaling paradigm has shaped decades of waveform design, ranging from single-carrier to multi-carrier orthogonal frequency-division multiplexing (OFDM) transmission and its diverse relatives.
However, wireless channels typically destroy orthogonality at the receiver side, 
which makes Nyquist signaling futile to a degree. It is also worth mentioning that the 2G Global System of Mobile (GSM) communications opted for deliberately introducing controlled ISI by harnessing partial-response GMSK signaling for spectral compactness, which required a channel equalizer.

Similarly, faster-than-Nyquist (FTN) signaling 
deliberately transmits symbols at a rate higher than the Nyquist rate, thereby also introducing controlled ISI~\cite{FTNMAZO}.
Therefore, FTN signaling can be considered as a special type of partial response signaling (PRS) that imposes intentional ISI on the transmitted signal. Note again that PRS aims for improving the bandwidth efficiency by introducing ISI for the sake of shaping the power spectral density (PSD) of the transmitted signal according to the bandwidth constraint. 
However, this is different from FTN signaling, whose PSD is usually not subject to optimization and is solely dependent on the choice of the shaping pulse adopted. 
In fact, FTN signaling directly increases the bit rate without directly altering the PSD or expanding the signal bandwidth, hence, leading to an improved constrained capacity, as it will be shown later in this paper.

\subsection{History of FTN Signaling}
Indeed, FTN technology has garnered increasing attention in recent years, but its concept germinated in the previous century. Notably, Shannon's seminal 1948 paper already hinted at the principles underlying FTN transmission. In this work, Shannon considered a scenario where the channel bandwidth does not match with the signal bandwidth. The following excerpt from his original text captures this idea:

``\textit{Let these functions be passed through an ideal filter with a triangular transfer characteristic. The gain is to be unity at frequency $0$ and decline linearly down to gain $0$ at frequency $W$... First we note that a pulse $\frac{\sin 2 \pi W t}{2 \pi W t}$ going into the filter...}''

Observe that the signal bandwidth{\footnote{Shannon did not deliberately consider negative frequencies.}} considered was $W$, while the bandwidth of the transmitted signal was 
$2W$. Essentially, this is FTN signaling with a symbol rate twice the channel bandwidth. Although Shannon considered this signaling approach, his paper did not elaborate further on the relationship between channel capacity and this mismatch.

While most FTN literature recognizes Mazo as the inventor of FTN signaling, the term ``faster-than-Nyquist'' firstly appeared 
as early as 1970~\cite{Ishihara2021evolution}.
Shortly thereafter, in 1975, Mazo formally defined and analyzed FTN signaling~\cite{FTNMAZO}. 
Specifically, Mazo focused on analyzing the relationship between the minimum distance of FTN signals and the symbol rate under sinc pulse shaping. He found that increasing the symbol rate to a certain extent does not reduce the minimum Euclidean distance. 
The constrained capacity of FTN signaling was studied by Rusek and Anderson in~\cite{rusek2009constrained}. Their results showed that increasing the rate further beyond a certain symbol rate,  does not increase channel capacity, and this threshold is determined by the bandwidth of the shaping pulse. In the 2018 Shannon Lecture, Shannon Award recipient Gottfried Ungerboeck also discussed FTN-related capacity issues, further sparking academic interest in FTN transmission. 



In contrast to~\cite{Ishihara2021evolution}, this article provides an accessible non-mathematical overview of its principles and applications, highlights advanced designs and potential solutions, and explores promising directions for future research. We commence from the
comparison of Nyquist and FTN signaling
in terms of their TD (TD) and frequency-domain (FD) properties. 
Then, we
show that the unique FD properties of FTN signaling can in fact improve the constrained capacity of the system. Specifically, our capacity analysis is based on the properties of folded-spectrum, whose connection to the symbol rate is also highlighted. Furthermore, the equalization of FTN signaling is discussed in detail, including both TD and FD equalization schemes. Based on these schemes, we present some resent results on coded FTN signaling, where promising performance is observed that exceeds the limit of constrained-capacity of Nyquist signaling.
The novel application of FTN signaling for sensing is also highlighted in the paper. Our results show that sensing with random FTN communication signals can avoid the undesired peaks in the ambiguity function, which leads to an improved sensing performance. Finally, some potential future research directions of FTN signaling are provided.

\section{From Nyquist Signaling to FTN Signaling} \label{sec:nyqtoFTN}
\begin{figure*}[pt]
\centering
\includegraphics[width=0.8\textwidth]{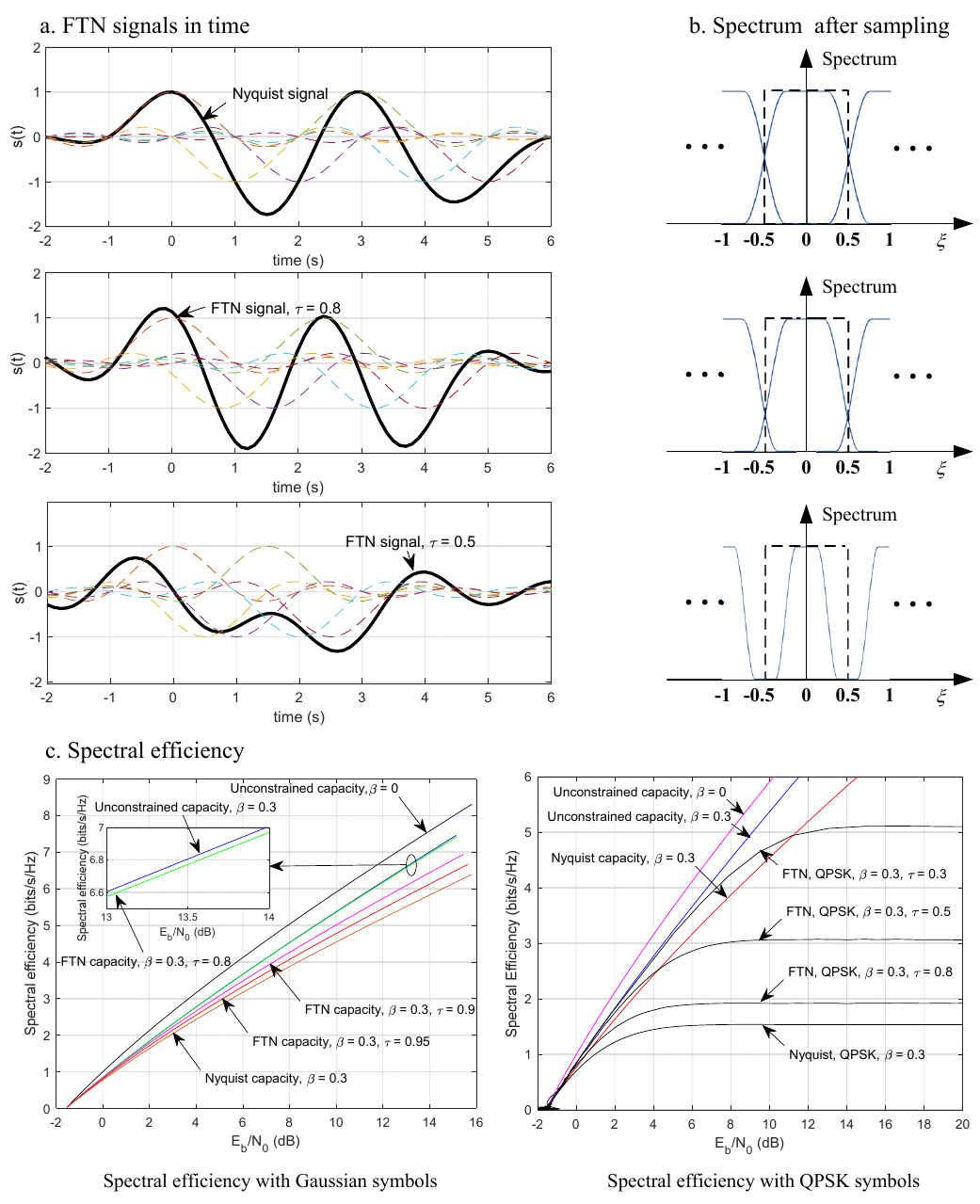}
	\caption{The TD FTN signals, the corresponding spectrum after sampling, and the spectral efficiencies of both Gaussian and non-Gaussian constellations, where the capacity results are derived based on independent and identically distributed symbols. }\label{FTN_diagram}
\end{figure*}

In this paper, we consider a general signaling format containing linearly modulated symbols of the form
\begin{align}
    s\left(t\right)=\sum_{n=1}^{N} x_n h\left(t-n \tau T\right),
    \label{general_signal}
\end{align}
where $x_n$ is the $n$-th entry of the transmitted symbol vector $\bf x$ of length $N$, $h\left(t\right)$ is a continuous energy-normalized TD pulse carrying the transmitted symbols, and $\tau T$ is the symbol duration. Particularly, $\tau \in \left(0,1\right]$ in~\eqref{general_signal} is the ``acceleration factor'' that controls the time separation between adjacent transmitted symbols. When $\tau=1$,~\eqref{general_signal} describes Nyquist signaling, while $\tau<1$,~\eqref{general_signal} represents FTN signaling. Some plots of signal with and without FTN appear in Fig.~\ref{FTN_diagram}.


After passing through an additive white Gaussian noise (AWGN) channel, the received signal is typically matched-filtered with respect to $h\left(t\right)$ and then sampled at the symbol rate $\tau T$. These sampled outputs form the effective channel observations for detecting the transmitted symbol vector $\bf x$.
Note that the continuous time signal in~\eqref{general_signal} has a bandlimited spectrum, whose shape is proportional to $|H\left(f\right)|^2$, i.e., the spectrum (squared Fourier transform) of the pulse $h\left(t\right)$. 
After FTN rate sampling, this spectrum is periodically extended according to the sampling rate $1/\tau T$. 
The spectrum after sampling is demonstrated also in Fig.~\ref{FTN_diagram}.
It is convenient to treat the pulse shaping, matched-filtering, and symbol-rate sampling together as a combined module, which has a  
periodically extended spectrum with respect to $|H\left(f\right)|^2$ within the nominal bandwidth. This is commonly referred to as the ``folded-spectrum'' defined by~\cite{anderson2017bandwidth} 
\begin{align}
&|{H}_{\rm fold}\left(\xi\right)|^2\notag\\
\triangleq& \left\{
\begin{array}{cc}
    \sum\limits_{n =  - \infty }^\infty\left|H\left( \frac{\xi+n}{\tau T}\right)\right|^2, & {\rm if} \; \xi \in\left[-\frac{1}{2}, \frac{1}{2}\right],  \\
    \quad\quad\quad\quad\quad\quad \;\;\;\;\;0, & \quad\quad\quad\quad \;\ {\rm else} ,
\end{array}\right.
\end{align}
which is the superposition of shifted scaled versions of $|H\left(f\right)|^2$ centered at all integer discrete frequencies.
It is important to note that the frequency period in the folded-spectrum is the reciprocal of the symbol time $\tau T$. Therefore, depending on the pulse $h\left(t\right)$ and the symbol duration $\tau T$ adopted, the shape of the folded-spectrum can change, affecting the communication performance.

An important aspect related to the folded-spectrum is the \textbf{symbol orthogonality}. According to the zero-ISI Nyquist theorem, the received symbols will  suffer from no ISI, if the folded-spectrum corresponds to a constant value. Such signaling is therefore referred to as Nyquist signaling, where simple symbol-by-symbol detection is sufficient thanks to symbol orthogonality. In other words, symbol orthogonality directly impacts both the performance and complexity of detection, which are critical aspects in practical communication systems.

Another important aspect related to the folded-spectrum is \textbf{spectral aliasing}. This 
occurs when the pulse spectrum 
$|H\left(f\right)|^2$ is not equal to the folded-spectrum
$|{H}_{\rm fold}\left(\xi\right)|^2$ in value for $\xi\in\left[-\frac{1}{2}, \frac{1}{2}\right]$~\cite{anderson2017bandwidth}. 
From an information-theoretical point of view, the level of spectral aliasing affects the degrees of freedom (DoFs) of the transmission, which influences the achievable communication rate.

\subsection{Nyquist Signaling} 
Most of the wireless waveforms are designed for Nyquist signaling.
A general Nyquist signaling can be represented by~\eqref{general_signal} with $\tau =1$, when $h\left(t\right)$ is a $T$-orthogonal pulse.  
The $T$-orthogonality of the pulse ensures that the zero-ISI Nyquist theorem is satisfied, thereby eliminating ISI among the transmitted symbols.

However, Nyquist signaling can suffer from spectral aliasing when $h\left(t\right)$ is not a perfect sinc pulse. Note that the minimum (two-sided) bandwidth requirement for a $T$-orthogonal pulse is $\frac{1}{T}$, which can only be achieved by the sinc pulse. Unfortunately, the ideal sinc pulse is not realizable in practice, because it has a perfect low-pass spectrum; therefore, the family of root-raised cosine (RRC) pulses are usually applied, which can have the same TD zero-crossings for retaining no ISI.
The bandwidth of a RRC pulse is given by $W=\frac{1+\beta}{T}$, with an excess bandwidth controlled by the roll-off factor $\beta$ with $0 \le\beta \le 1$, which reduces to the sinc pulse for $\beta=0$. As a result, Nyquist signaling suffers from spectral aliasing when $\beta>0$. An illustration of Nyquist signaling with RRC pulses is presented in the upper part of Fig.~\ref{FTN_diagram}.a, while its spectrum after sampling is also highlighted in Fig.~\ref{FTN_diagram}.b. As indicated by the figure, the spectrum appears to have severe overlapping between adjacent replicas. According to the properties of RRC pulses, the overlapped spectrum appears as a 
rectangular shape, which satisfies the zero-ISI Nyquist criterion but introduces spectral aliasing.

\subsection{FTN Signaling} 
Again, a general FTN signaling waveform can be represented by~\eqref{general_signal} with $\tau <1$. Since FTN signaling intentionally introduces non-orthogonality among the transmitted symbols, the application of $T$-orthogonal pulses is not necessary.
Although non-orthogonal FTN signaling signaling potentially requires high complexity equalizers for ISI-cancellation, its appealing spectral efficiency makes it stand out.

The TD FTN signals are presented in Fig.~\ref{FTN_diagram}.a, while the corresponding spectrum after receiver sampling is provided in Fig.~\ref{FTN_diagram}.b.
For illustration, we
consider two FTN cases, namely $\tau=0.8$ and $\tau=0.5$, using an RRC pulse having $\beta=0.5$. As $\tau$ decreases, the bandlimited signaling pulses become more densely packed in time, leading to tighter temporal overlap. The corresponding spectrum after sampling exhibits distinct characteristics for different values of $\tau$. For the moderate acceleration factor of $\tau=0.8$, the corresponding spectrum has
slight overlapping between adjacent spectrum replicas, causing spectral aliasing. By contrast, for a more aggressive acceleration factor of $\tau=0.5$, the spectrum closely resembles the original spectral shape, indicating that spectral aliasing is effectively avoided. However, it is important to note that when $\tau < \frac{1}{WT}$, the resultant spectrum may exhibit spectral nulls, as illustrated in the figure. These spectral nulls can further complicate the equalization in FTN systems and we will revisit this issue in Section \ref{sec:ftneq}.

\section{Constrained Capacity of FTN Signaling} \label{sec:ftncap}
Channel capacity is a fundamental characteristic of communication systems.
In the context of FTN signaling, its capacity is often investigated based on the discrete-time system model that considers the receiver FTN rate sampling, i.e., a capacity constrained by FTN rate transmission and sampling. Notice that this constrained capacity can be different from the continuous-time capacity of the signal~\eqref{general_signal}, due to the potential spectral aliasing caused by receiver sampling.

The constrained capacity of FTN signaling was investigated in \cite{rusek2009constrained}, where symbols are assumed to be independent and identically distributed and the receiver employs matched filtering followed by sampling at a rate of $1/\tau T$.
Let $S_x\left(\xi\right)$ be the PSD of the discrete information symbols. Then, the constrained capacity of signaling according to~\eqref{general_signal} is given by
\begin{align}
    C=\frac{1}{\tau T}\int_{-\frac{1}{2}}^{\frac{1}{2}} \log_2 \left(1+\frac{S_x\left(\xi\right)}{N_0}|H_{\rm fold} \left(\xi\right)|^2\right) {\rm d} \xi \; {\rm bits/s},
    \label{Shannon_capacity}
\end{align}
where $N_0$ denotes the one-sided PSD of the additive white Gaussian noise (AWGN) process.
Notice that the folded-spectrum $|H_{\rm fold} \left(\xi\right)|^2$ in~\eqref{Shannon_capacity} is closely related to both the pulse spectrum  $|H\left(f\right)|^2$ and the acceleration factor $\tau$. Hence, one can optimize $|H\left(f\right)|^2$ and $\tau$ for improving the constrained capacity. 
It can be observed from~\eqref{Shannon_capacity} that the bandwidth of the folded-spectrum corresponds to the maximum achievable DoF. Notably, when $\tau <\frac{1}{WT}$, spectral aliasing is avoided, and the folded-spectrum preserves the original shape of $|H\left(f\right)|^2$. In this case, the constrained capacity aligns with the classical Shannon capacity associated with the pulse spectrum, achieving the maximum rate for independent and identically distributed symbols. 

This constrained capacity framework offers interesting insights into the performance difference between FTN and Nyquist transmissions. In Nyquist signaling, the zero-ISI Nyquist condition requires the folded-spectrum to be flat. However, this condition can only be fulfilled via spectral aliasing, when the shaping pulse is non-sinc. 
Although sinc pulses are theoretically optimal for achieving the highest spectral efficiency over AWGN channels, they cannot be used in real-world systems. In practice, all systems use non-sinc pulses, which lead to excess bandwidth and a corresponding spectral efficiency erosion. Given this practical limitation, FTN signaling offers an advantage over traditional Nyquist signaling by making more efficient use of the available bandwidth. Indeed, it has been shown in~\cite{rusek2009constrained} that FTN signaling has a higher constrained capacity than the Nyquist signaling, when non-sinc pulses are applied.

It is important to note that the preceding discussion focuses on constrained capacity with independent and identically distributed inputs. 
To further improve the spectral efficiency, one can optimize the input spectrum. It is known that a flat (constant) PSD over the assigned signal bandwidth is optimal for transmissions over AWGN channels. Therefore, the optimal assigned input spectrum is the one that ensures the effective frequency spectrum $S_x\left(\xi\right)|H_{\rm fold} \left({\xi}\right)|^2$ is rectangular in shape, corresponding to uniform power allocation across the occupied frequency band. In this case, the FTN signal effectively reduces to a Nyquist signal using a sinc pulse with an increased bandwidth, yet, has a higher spectral efficiency than~\eqref{Shannon_capacity}. 

The capacity analysis of FTN signaling has stimulated valuable discussions related to holistic communication system design. For instance, Kim~\cite{kim2016properties} studied the benefits of receiver oversampling in Nyquist signaling with non-sinc pulses and concluded that this method fails to improve achievable rates, owing to the limited rank (dimensionality) of the effective channel matrix. Furthermore, Yoo and Cho~\cite{Geon2010asymptotic} showed the asymptotical optimality of FTN signaling along with non-Gaussian constellations. By using appropriate approximations, they demonstrated that the achievable FTN rate of non-Gaussian constellations approaches to the constrained capacity in~\eqref{Shannon_capacity} with a reduced $\tau$.
Intuitively, this is because
the increased ISI turns the effective channel input into resemble near-Gaussian random variable, even when the underlying modulation (e.g., BPSK) is non-Gaussian, as $\tau$ approaches zero.

The spectral efficiency of FTN signaling with different acceleration factors is plotted in Fig.~\ref{FTN_diagram}.c, for both Gaussian (left) and QPSK constellations (right). In the figure, the horizontal axis is the average bit energy to noise ratio, where $E_{\rm b} = E_{\rm s}/R$, with $E_{\rm s}$ is the average baseband symbol energy, and $R$ being the achievable rate. 
The results recorded for Gaussian constellations are computed according to~\eqref{Shannon_capacity}, where  
we observe that the spectral efficiency improves as $\tau$ decreases, up to the point where $\tau <\frac{1}{WT}$. In this regime, the spectral efficiency of FTN has a near-constant gap with respect to the ideal Nyquist signaling with sinc pulse, i.e., $\beta=0$, further validating the DoF optimality of FTN signaling.
The results plotted for QPSK constellations are obtained using the Arnold-Loeliger algorithm~\cite{rusek2009constrained} by considering a sufficiently large number of ISI taps corresponding to $\tau$. Observe that the achievable rate increases upon decreasing $\tau$, and this improvement persists even when $\tau < \frac{1}{WT}\approx 0.769$, which validates that 
reducing $\tau$ can provide shaping gains for non-Gaussian constellations.

Finally, it is worth emphasizing that the above discussions assume optimal detection, which is rarely feasible in practice due to its prohibitive complexity. In reality, reduced-complexity detection is a critical enabler for practical FTN systems. This important topic will be discussed in detail in the next section.

\section{FTN Equalization} \label{sec:ftneq}
Equalizing FTN signals can be challenging due to severe ISI, especially when the acceleration factor $\tau$ is low. 
Note that sampling at the FTN rate introduces \textit{colored noise} at the receiver,
making the FTN equalization problem more challenging than traditional equalization for ISI channels.
Most existing research on FTN equalization can be broadly categorized into either TD or frequency domain (FD) approaches, which will be discussed in this section.
Before delving into FTN equalization, we first present the important related concept of Mazo limit, which was the primary motivation for considering FTN signaling back in the 1970s.

\subsection{Mazo Limit} 
The Mazo limit provides insights into FTN detection performance. In his work~\cite{FTNMAZO}, Mazo showed that for uncoded BPSK signaling employing sinc pulse shaping, the minimum Euclidean distance $d_{\min}^2$ between symbol sequences remains unchanged even as the symbol rate increases up to a critical threshold known as the ``Mazo limit''. This finding implies that FTN signaling can support increased symbol rates without substantially degrading the error performance, provided that maximum-likelihood sequence estimation (MLSE) is used.
Subsequent research confirmed the existence of Mazo limits for other practical pulse shapes, including the RRC pulses. For instance, the Mazo limit is $\tau=0.802$ for the sinc pulse, $\tau=0.779$ for an RRC pulse with roll-off factor $\beta=0.1$, and decreases further to $\tau=0.703$ for $\beta=0.3$, reflecting the intuitive observation that the Mazo limit decreases with a larger excess bandwidth.
The physical explanation behind the Mazo limit is not mysterious. In Nyquist signaling, the error event corresponding to $d_{\min}^2$ is known to be the \textit{antipodal event} that contains only a single symbol error. In FTN signaling, as $\tau$ is reduced, other error events  arise that match the Euclidean distance of the antipodal event. The specific value of $\tau$, where this occurs, defines the Mazo limit~\cite{anderson2017bandwidth}.

The confirmation of the Mazo limit suggests that, under optimal MLSE detector, FTN signaling does not experience significant error performance degradation relative to Nyquist signaling, as long as 
$\tau$ remains above the Mazo limit. In practical terms, this means that FTN signaling employing sinc pulse shaping can transmit approximately $25\%$ more BPSK symbols over the same bandwidth without compromising reliability.
However, it is crucial to recognize that the Mazo limit applies only to uncoded FTN signaling under optimal MLSE detection. In practice, as $\tau$ becomes smaller, the effective channel memory grows significantly, making MLSE computationally demanding. Therefore, the Mazo limit provides only limited practical guidance. Encouragingly, recent studies have shown that coded FTN signaling using reduced-complexity detectors can achieve excellent performance even at acceleration factors well below the Mazo limit. These advances suggest that practical FTN systems can operate reliably under aggressive symbol packing, and this topic will be further explored in Section~\ref{sec:codedftn} of this paper.

\subsection{Time-Domain Equalization} 
TD equalizers based on ISI graph models, e.g., trellises, can achieve good error performance for FTN detection, albeit their complexity typically escalates exponentially with the channel memory. In the context of FTN signaling, such models are primarily derived from three observation frameworks, namely the Forney observation model, the Ungerboeck observation model, and the orthogonal basis model~\cite{Prlja2012MBCJR}.

The Forney observation model applies a whitened matched filter (WMF) at the receiver, ensuring that the resultant noise becomes white. Although it has the benefit of analytical convenience, it can suffer from numerical instability when $\tau < \frac{1}{WT}$, due to having spectral nulls in the folded-spectrum (see Fig.\ref{FTN_diagram}.b). By contrast, the orthogonal basis model is entirely applicable, when such spectral nulls exist~\cite{anderson2017bandwidth}. This model represents the transmit shaping pulse $h\left(t\right)$ using a sequence of wider-band orthonormal pulses 
$\phi \left(t\right)$, which are $\tau T$-orthogonal~\cite{Prlja2012MBCJR}, thus they only exist when $\frac{1}{\tau T} > W$.
In the orthogonal basis model, the receive shaping pulse is $\phi \left(t\right)$ instead of $h\left(t\right)$, so that the resultant noise samples remain white.
In contrast, the Ungerboeck observation model avoids any noise-whitening operations and operates directly in the presence of colored noise, which is also widely harnessed for FTN equalization.

For challenging long memories, a complete BCJR algorithm based on the FTN trellis usually imposes excessive complexity. To address this, most BCJR-based FTN equalizers adopt either \textit{reduce-search} or \textit{reduce-size} strategies.
Typical reduce-search strategies, such as $M$-BCJR and T-BCJR, compute sequence likelihoods over a subset of the ISI trellis. At each trellis depth, a candidate set of likely states is dynamically selected based on likelihood metrics. This is straightforward under the Forney and orthogonal basis models, where each state probability is directly known from the path metric. However, for the Ungerboeck model, where such metrics lack a direct probabilistic interpretation, conceiving reduced-search is more challenging.  To this end, a modified $M$-BCJR algorithm was proposed in~\cite{Shuangyang2018Ungerboeck}, which uses breadth-first search across multiple future trellis depths to select near-optimal state subsets in a near maximum-likelihood sense.
On the other hand, the reduce-size BCJR operates on a simplified trellis by intentionally ignoring some ISI taps. While this reduces complexity, it may lead to performance erosion due to model mismatch.

For both reduce-search and reduce-size equalization, employing an optimized filter after matched filtering can significantly improve error performance.
For reduced-search BCJR, minimum-phase ISI models are typically preferred. By concentrating energy toward the early part of the ISI response, the trellis state selection becomes more accurate. Minimum-phase models can be constructed by applying all-pass filters to the received signal. In~\cite{Prlja2012MBCJR}, a super minimum-phase model was constructed for FTN via such filtering, allowing $M$-BCJR to operate with $2$ to $4$ times fewer trellis states compared to the unfiltered model.
For reduced-size BCJR, typically channel shortening filters are applied, which are designed for concentrating ISI energy into a reduced number of taps. These filters aim for reducing the effective channel memory so that neglecting the residual ISI only has minimal impact on error performance. Since FTN signaling typically has long but weak ISI tails, channel shortening has demonstrated promising error performance.

\subsection{Frequency-Domain Equalization} 

Frequency-domain equalization (FDE) exploits the inherent Toeplitz structure of the ISI channel matrix in FTN signaling. By appending a sufficiently long cyclic prefix (CP) to the transmitted FTN symbols, the Toeplitz matrix can be approximated as being circulant, hence allowing diagonalization via the discrete Fourier transform (DFT). This allows reduced-complexity FD equalization. The application of FDE to FTN signals was first proposed in~\cite{Sugiura2013freq}, demonstrating promising error performance for relatively high acceleration factors $\tau$.

FDE is particularly attractive for FTN detection due to its low computational complexity, hence it has been adopted in diverse FTN studies. However, it also has its limitations. Firstly, the use of the CP imposes a throughput erosion compared to TD equalization, which does not require CP. More critically, the performance of FDE degrades as 
$\tau$ decreases. According to~\cite{kim2016properties}, the eigenvalues of the FTN channel matrix correspond to samples of the folded-spectrum, taken at intervals of $\frac{1}{N \tau T}$, where $N$ denotes the number of transmitted symbols. Consequently, when $\tau <\frac{1}{WT}$, the folded-spectrum contains spectral nulls. In such cases, the corresponding subchannels carry no reliable information, hence resulting in severe performance degradation.
To address this issue, FD power allocation has been considered. An eigenvalue decomposition (EVD)-based precoding scheme was introduced in~\cite{Ishihara2022Eigen}, where a power allocation matrix is applied across the frequency bins to address the aforementioned problem. Specifically, when this power allocation follows the water-filling principle, the scheme can theoretically achieve the unconstrained channel capacity~\cite{zhang2025capacity}. As discussed in Section~\ref{sec:ftncap}, this approach is functionally equivalent to transmitting Nyquist signals using a sinc shaping pulse with an increased bandwidth.

It should be highlighted that FDE enjoys appealing advantages for detecting FTN signals transmitted over non-AWGN channels. For example, when transmitting over the frequency-selective channel, FTN systems can suffer from severe ISI caused by both the FTN transmission and the channel dispersion in time. In this case, equalization in the TD can be very challenging since the required complexity can be prohibitive. In contrast, the required complexity of FDE remains roughly unchanged since it complexity does not related to the length of effective ISI taps, which makes it an attractive solution.

\section{Coded FTN Signaling} \label{sec:codedftn} 
The adoption of FTN signaling opens new avenues for innovative channel coding strategies. As discussed in Section \ref{sec:ftncap}, FTN signaling offers enhanced constrained capacity compared to Nyquist signaling. Therefore, with appropriately designed channel codes, coded FTN signaling has the potential to achieve superior performance relative to its Nyquist counterparts.

Most coded FTN schemes can be viewed through the lens of pulse-shaped concatenated coding, where FTN modulation serves as an ``inner code'', concatenated with an outer code via an interleaver. To decode such systems, turbo equalization is typically employed, iteratively exchanging soft information between the FTN equalizer and the outer decoder. In this framework, powerful code design tools such as extrinsic information transfer (EXIT) charts become applicable. For instance,~\cite{Yang2025analysis} proposed an improved 5G low-density parity check (LDPC) code design for FTN signaling. Specifically, two rule-of-thumb code design  criteria were revealed: 1) no information bits should be punctured; 2) columns having high weights should be removed on the base graph. 
Intuitively, these principles aim for preventing the decoder from becoming overconfident during the early iterations of turbo equalization, which is crucial since the initial soft information gleaned from the FTN equalizer is generally less reliable.
Hence, outer codes that generate reliable extrinsic information at low signal-to-noise ratios (SNRs) are generally preferred in FTN systems.

An alternative approach is presented in~\cite{Li2020code}, which employs a specialized class of convolutional codes known as output-retainable codes, operating without any interleaving. In this design, the convolutional code trellis and the FTN trellis partially overlap, 
hence the code can be optimized by minimizing the effective Euclidean distance among different codewords.
A key advantage of output-retainable codes is that certain future code symbols can be inferred from the current code trellis state, allowing a reduced-complexity BCJR algorithm to be applied directly to the code trellis. This scheme inherits its motivation from the channel shortening receiver and it is therefore termed as a ``code based channel shortening'' scheme. Furthermore, 
this convolutional coded FTN scheme can also be concatenated with another code for further improving the error performance~\cite{Li2020code}.

\begin{figure}[pt]
\centering
\includegraphics[width=0.45\textwidth]{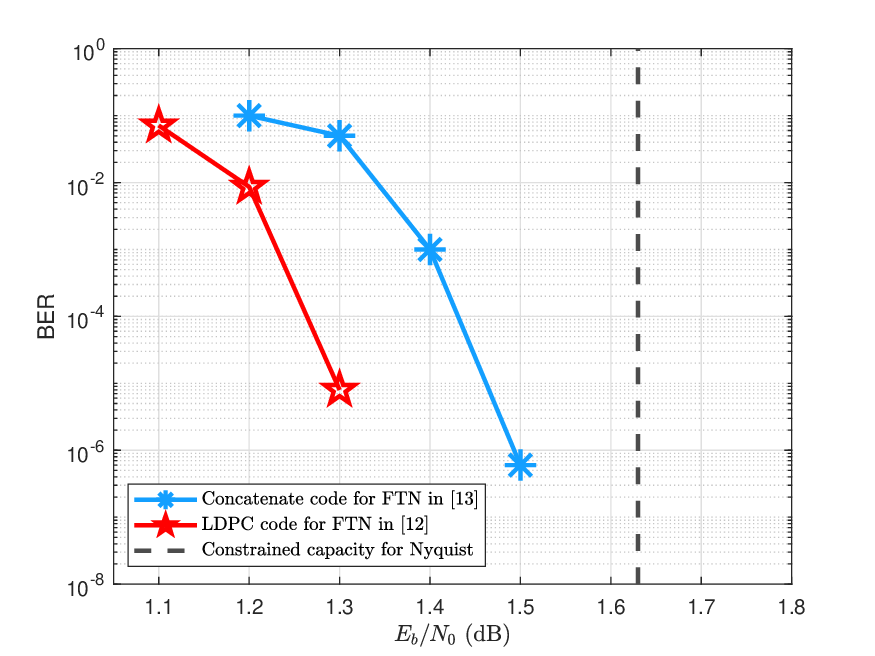}
	\caption{BER performance of coded FTN signaling, compared to the constrained capacity of Nyquist signaling. }\label{FTN_code}
\end{figure}

Fig.~\ref{FTN_code} presents the BER performance of coded FTN signaling with $\tau= 2/3$, comparing LDPC coded systems~\cite{Yang2025analysis} with concatenated convolutionally coded systems~\cite{Li2020code}, where the codeword length is roughly $64,800$ and the code rate is $0.5$. As shown in Fig.~\ref{FTN_code}, both approaches operate reliabily below the constrained capacity of Nyquist signaling. Additionally, the optimized LDPC coded FTN signaling achieves a performance gain of approximately $0.15$ dB over the concatenated scheme, albeit at the cost of increased detection and decoding complexity.

It should be noted that the above results also demonstrate the practical advantages of coded FTN signaling in comparison to the Nyquist counterpart.
Specifically, with FTN signaling, the communication signal with low code rate can be transmitted much faster. Effectively, this can significantly boost the spectral efficiency without increasing the required time frequency resources compared to Nyquist signaling.
Therefore, coded FTN signaling allows flexible adjustment on the symbol rate and code rate, thereby achieving a good tradeoff between the equalization complexity and error-correction ability.
This is particularly beneficial for practical transmission in the mid-to-low SNR regime, where the error performance is primarily constrained by the error-correction ability of the code. As a result, it is expected that the coded FTN signaling with low code rate and relatively high acceleration factor can outperform coded Nyquist signaling with a high code rate, even when a reduced-complexity equalizer is adopted, as demonstrated in the above results. This makes coded FTN  attractive also for energy-efficient practical deployments.

\section{Integrated Sensing and Communications using FTN Signals}

ISAC services have been identified as one of the six primary use cases for 6G. As such, the study of FTN signaling in ISAC systems is highly relevant to future standardization efforts. In particular, next-generation wireless networks require extremely high data rates, which naturally shift ISAC operations toward a communication-centric mode. In this context, it is important to evaluate the achievable sensing performance, when FTN signals are used primarily for data transmission.

A common approach to analyze the sensing capability of communication-centric ISAC signals is through the ambiguity function, which characterizes the sensitivity of the transmitted signal to different delay and Doppler offsets, when a matched-filtering based sensing receiver is applied.
However, because communication signals are inherently random, the ambiguity function has to be evaluated in an expected sense. 
The expected squared ambiguity function of FTN signals was derived in~\cite{FTN_ISAC2025}, where the randomness of the communication symbols was exploited.
The result confirms that the sensing performance of FTN signaling critically depends on the relationship between signal bandwidth and symbol rate. Specifically, when the signal bandwidth is lower than the symbol rate, spectral aliasing occurs, as discussed in Section~\ref{sec:nyqtoFTN}. This aliasing introduces fluctuations in the expected squared ambiguity function along the delay axis. More significantly, aliasing leads to undesired peaks along the Doppler dimension, potentially causing ambiguities in Doppler sensing. It was shown that these undesired peaks appear at integer multiples $\frac{1}{\tau T}$, which can be avoided when the bandwidth of the shaping pulse is lower than $\frac{1}{\tau T}$. 
This aligns with the motivation of applying FTN signaling for communications and verifies the advantages of FTN signaling for sensing.

\begin{figure}[pt]
\centering
\includegraphics[width=0.45\textwidth]{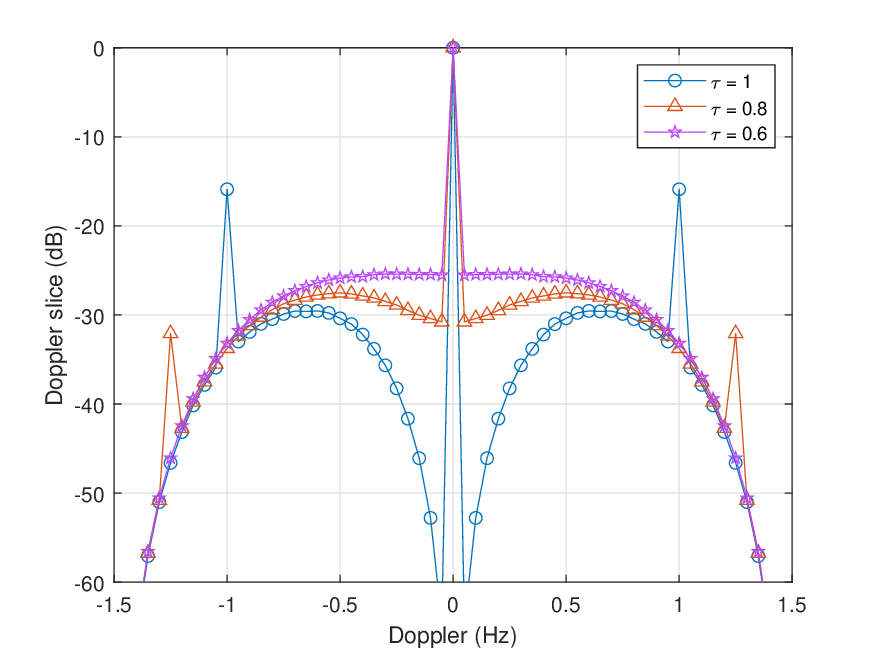}
	\caption{Comparison of the Doppler slices of the expected squared ambiguity functions with $\tau=1$, $\tau=0.8$, and $\tau=0.6$, where the RRC pulse with $\beta = 0.5$ is adopted as the shaping pulse.
    }\label{AF_nu_xi_compare}
\end{figure}

\begin{figure}[pt]
\centering
\includegraphics[width=0.45\textwidth]{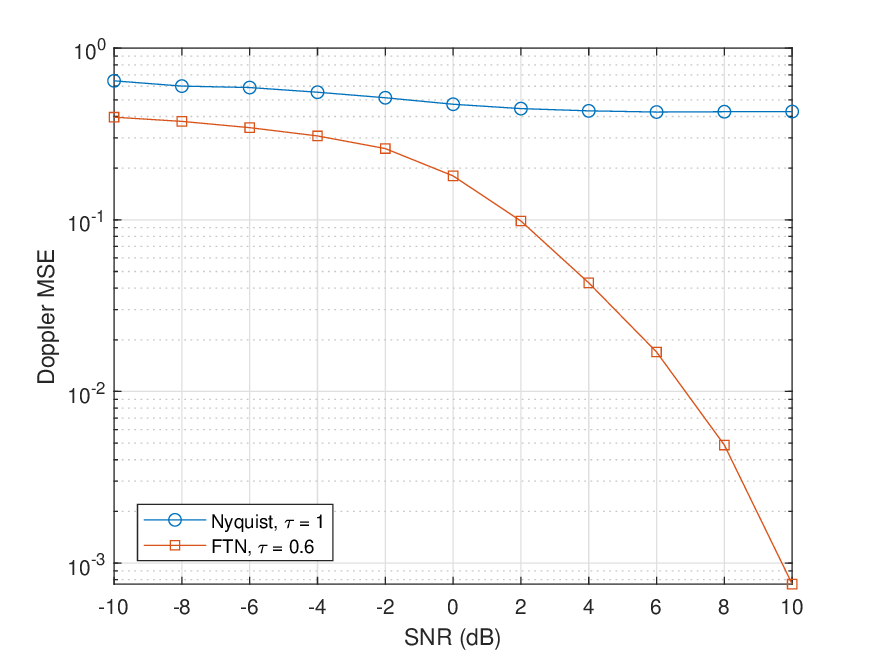}
	\caption{Doppler sensing performance of both Nyquist and FTN signals with $\tau=0.6$, where the RRC pulse with $\beta = 0.5$ is adopted as the shaping pulse. Here, one strong target and one weak target ($15\%$ of the reflectivity strength) are considered with normalized Doppler at $0.5$ Hz and $-0.4$ Hz.  }\label{Actual_Sensing_performance}
\end{figure}

We illustrate the Doppler slice of the expected squared ambiguity functions of FTN signals having different acceleration factors in Fig.~\ref{AF_nu_xi_compare}, where an RRC pulse with roll-off factor $\beta = 0.5$ is used for pulse shaping. As shown in the figure, undesired peaks appear in the ambiguity functions for $\tau =1$ and $\tau=0.8$, whereas no such peaks are observed for the $\tau =0.6$ case. This behavior is consistent with the discussion above.
To further evaluate the impact of FTN signaling on practical sensing performance, Fig.~\ref{Actual_Sensing_performance} presents the Doppler sensing results for both Nyquist and FTN signals using the same RRC pulse ($\beta = 0.5$) based on maximum-likelihood estimation. For FTN signaling, an acceleration factor of $\tau=0.6$ is used. The sensing scenario considered involves a strong and a weak target (with 
$15\%$ of the reflectivity strength of the strong target), whose  normalized Doppler shifts are $0.5$ Hz and $-0.4$ Hz, respectively.
As demonstrated in the figure, the Nyquist signal fails to provide accurate Doppler sensing performance, while the FTN signal excels, which confirms the advantage of FTN signaling in communication-centric ISAC systems.

We highlight that although FTN signaling can yield improved sensing performance theoretically, practical sensing receiver design tailored to FTN signaling have not been fully studied yet. 
Conventional sensing receivers based on matched-filtering or maximum-likelihood estimation can be used but they may suffer from high complexity in practice. Therefore, new sensing receivers exploiting the unique properties of FTN signaling shall be developed in order to facilitate the sensing application using FTN signals.

\section{Future Directions}

\subsection{Pulse Design and Generalized FTN Signaling}
Most existing FTN literature has primarily focused on the use of 
$T$-orthogonal pulses. However, as previously discussed, this orthogonality constraint is not a strict requirement for FTN signaling. Hence, non-$T$-orthogonal pulses can also be considered. This framework opens up new research directions in pulse shape design, where the objective is to enhance communication performance, while meeting practical constraints on bandwidth and time duration.
A potential way to design pulses is to optimize the folded-spectrum for improving achievable rates.
Alternatively, pulse design can aim for maximizing the minimum Euclidean distance between codewords, thereby enhancing detection performance.
In addition, novel pulses may facilitate the FTN equalization with reduced-complexity, e.g., the pulse shaping considered in~\cite{Prlja2012MBCJR}, which is crucial for practical applications of FTN signaling and its standardization. Despite these opportunities, the design and analysis of generalized FTN signaling remain relatively underexplored and represent a valuable direction for future investigations.

\subsection{FTN Signaling in ``Alternative Domains''}
The FTN signaling discussed in this paper focuses on squeezing together consecutive TD pulses, which can be viewed as a direct extension of single-carrier Nyquist signaling. However, the fundamental concept of FTN, which is to deliberately violate orthogonality for enhancing spectral efficiency, can also be extended to alternative signal domains via non-unitary precoding.
For instance, spectrally efficient frequency division multiplexing (SEFDM)~\cite{xu2014spectrally} applies non-orthogonal symbol multiplexing in the FD to achieve higher spectral efficiency. More recently, the consideration of the symbol multiplexing in alternative domains, such as the delay-Doppler domain, and the affine Fourier domain, has also emerged. The extension of FTN signaling to these domains offers promising opportunities for advanced waveform design, enabling further potential gains in communication performance beyond conventional TD squeezing. Such FTN techniques can play an important role in future high-efficiency, application-specific communication systems.

\subsection{FTN Signaling with Interference Pre-cancellation}
Since the interference of FTN signaling is known at the transmitter, it is possible to apply pre-cancellation schemes to eliminate the ISI before the transmission. A possible solution for the pre-cancellation may be the Tomlinson-Harashima precoding (THP), where the QR decomposition can be firstly applied to the FTN ISI matrix followed by a symbol-wise interference cancellation according to the triangular matrix structure.
However, the FTN ISI matrix can be highly ill-conditioned due to the high symbol rate, which can cause error performance degradation.
Furthermore, THP is known to be suboptimal in terms of the achievable rate due to the ``shaping loss'' and ``modulo loss''. However, their impacts on FTN signaling is still unknown. Overall, the interference pre-cancellation for FTN signaling is still underdeveloped and can be a good solution for applications with strict receiver complexity constraints.

\subsection{FTN Signaling over Multipath Channels}
Although FTN equalization has been extensively studied over the years, it remains computationally demanding, particularly in multipath fading channels. In such environments, multipath propagation can further complicate the interference pattern and pose additional challenges for FTN receiver design.
Most conventional FTN equalizers have been developed for AWGN channels, and only a limited number of low-complexity solutions exist for FTN equalization in the presence of multipath. Other than the FDE discussed in Section IV-C, a promising direction is the use of iterative equalization techniques that partition the complex interference structure into several interconnected subproblems having reduced local interference. Employing message passing between these subcomponents may improve detection performance at moderate complexity.

\subsection{FTN Signaling with Finite Block-Length Transmission}
Given that FTN signaling can significantly extend the payload without requiring additional time-frequency resources, its application to finite block-length transmission is well-motivated. Most existing studies on FTN signaling focus on the infinite block-length regime, hence their results cannot be directly applied to the finite block-length setting.
Nonetheless, FTN signaling is expected to offer improved achievable rates compared to Nyquist signaling under identical time-frequency constraints. This partly due to its higher Shannon capacity, 
and partly because of the potentially reduced channel dispersion, which generally decreases with increasing frame length. Despite this promising potential, research on finite-length FTN signaling requires further in-depth investigations.

\subsection{FTN Signaling for Secure Transmission}
When the acceleration parameter $\tau$ is unknown at the receiver, reliable decoding of FTN signals becomes fundamentally problematic because the ISI matrix, whose entries depend on $\tau$ and the shaping pulse, cannot be specified accurately. Any mismatch in $\tau$ can distort the FTN ISI matrix, yielding model error that severely degrades equalization performance. In practice, the receiver must either adopt non-coherent detection, accepting the performance loss from discarding the precise channel structure, or jointly estimate 
$\tau$ together with data symbols. However, none of these schemes are easy in practice. Consequently, it is possible to apply FTN signaling for secure transmission, where the eavesdropper has no prior information on $\tau$ and consequently cannot reliably decode FTN signals. However, the discussion on FTN signaling for secure transmission is still in its infancy and more studies are needed.

\section{Conclusions}
FTN signaling constitutes a bandwidth-efficient design paradigm for next-generation wireless networks. 
We commenced by summarizing recent efforts towards the standardization of FTN signaling. Then, we provided an overview of the fundamental principles of FTN signaling, emphasizing its distinguishing features and advantages over conventional Nyquist signaling. Furthermore, we discussed emerging applications of FTN signaling in the context of ISAC. We further outlined several promising research directions, highlighting key challenges and opportunities for further exploration. It is hoped that this article will inspire continued research in this vibrant area and contribute to the development of next-generation communication systems.

\bibliographystyle{IEEEtran}
\bibliography{reference}

@article{rusek2009constrained,
  title={Constrained capacities for faster-than-{Nyquist} signaling},
  author={Rusek, Fredrik and Anderson, John B},
  journal={IEEE Trans. Inf. Theory},
  volume={55},
  number={2},
  pages={764--775},
  year={2009},
  month={Feb.},
  publisher={IEEE}
}

@ARTICLE{Li2020code,
  author={S. {Li} and J. {Yuan} and B. {Bai} and N. {Benvenuto}},
  journal={IEEE Trans. Commun.},
  title={Code-Based Channel Shortening for Faster-Than-{Nyquist} Signaling: Reduced-Complexity Detection and Code Design},
  year={2020},
  volume={68},
  number={7},
  month={Jul.},
  pages={3996-4011},}

@inproceedings{kim2016properties,
  title={Properties of {faster-than-Nyquist} channel matrices and folded-spectrum, and their applications},
  author={Kim, Yong Jin Daniel},
  booktitle={IEEE Wireless Commun. Net. Conf.},
  pages={1--7},
  year={2016},
}

@ARTICLE{FTNMAZO,
  author={Mazo, J. E.},
  journal={The Bell System Techn. J.}, 
  title={Faster-than-{Nyquist} signaling}, 
  year={1975},
  volume={54},
  number={8},
  pages={1451-1462},
month={Oct.},
  doi={10.1002/j.1538-7305.1975.tb02043.x}}

@ARTICLE{Shuangyang2018Ungerboeck,
  author={Li, Shuangyang and Bai, Baoming and Zhou, Jing and Chen, Peiyao and Yu, Zhongyang},
  journal={IEEE Trans. Commun.}, 
  title={Reduced-Complexity Equalization for Faster-Than-{Nyquist} Signaling: New Methods Based on {Ungerboeck} Observation Model}, 
  year={2018},
  volume={66},
  number={3},
month={Nov.},
  pages={1190-1204},
  doi={10.1109/TCOMM.2017.2774816}}

@ARTICLE{Prlja2012MBCJR,
  author={Prlja, Adnan and Anderson, John B.},
  journal={IEEE Trans. Commun.}, 
  title={Reduced-Complexity Receivers for Strongly Narrowband Intersymbol Interference Introduced by Faster-than-{Nyquist} Signaling}, 
  year={2012},
  volume={60},
  number={9},
  pages={2591-2601},
month={Sept.},
  doi={10.1109/TCOMM.2012.070912.110296}}

@ARTICLE{FTN_ISAC2025,
  author={Li, Shuangyang and Liu, Fan and Xiong, Yifeng and Yuan, Weijie and Bai, Baoming and Masouros, Christos and Caire, Giuseppe},
  journal={IEEE Trans. Signal Process.}, 
  title={Faster-than-{Nyquist} Signaling is Good for Single-Carrier {ISAC}: An Analytical Study}, 
  year={2025},
  volume={73},
  number={},
  pages={3203-3219}}

@article{zhang2025capacity,
  title={Capacity and {IAPR} Analysis for {MIMO} Faster-than-{N}yquist Signaling with Small Acceleration Factors},
  author={Zhang, Zichao and Yuksel, Melda and Guvensen, Gokhan M and Yanikomeroglu, Halim},
  journal={{\emph to appear in} IEEE Trans. Wireless Commun.},
  year={2025}
}

@book{anderson2017bandwidth,
  title={Bandwidth Efficient Coding},
  author={Anderson, John B},
  year={2017},
  publisher={John Wiley \& Sons}
}

@ARTICLE{Geon2010asymptotic,
  author={Yoo, Young Geon and Cho, Joon Ho},
  journal={IEEE Commun. Lett.}, 
  title={ Asymptotic Optimality of Binary Faster-than-{Nyquist} Signaling}, 
  year={2010},
  volume={14},
  number={9},
  pages={788-790},
  month={Sept.},
  doi={10.1109/LCOMM.2010.072910.100499}}

@ARTICLE{Sugiura2013freq,
  author={Sugiura, Shinya},
  journal={IEEE Wireless Commun. Lett.}, 
  title={Frequency-Domain Equalization of Faster-than-{Nyquist} Signaling}, 
  year={2013},
  volume={2},
  number={5},
  pages={555-558},
month={Aug.},
  doi={10.1109/WCL.2013.072313.130408}}

@ARTICLE{Ishihara2022Eigen,
  author={Ishihara, Takumi and Sugiura, Shinya},
  journal={IEEE Trans. Wireless Commun.}, 
  title={Eigendecomposition-Precoded Faster-Than-{Nyquist} Signaling With Optimal Power Allocation in Frequency-Selective Fading Channels}, 
  year={2022},
  volume={21},
  number={3},
  pages={1681-1693},
month={Mar.},
  doi={10.1109/TWC.2021.3106098}}

@INPROCEEDINGS{Yang2025analysis,
  author={Yang, J. and Wang, Q. and Li, S. and Peng, K. and Ma, X. and Bai, B. and Caire, G.},
  booktitle={ IEEE Int. Symp. Inf. Theory}, 
  title={Analysis and Design of Improved {5G} {LDPC} Codes for Faster-Than-{Nyquist} Signaling}, 
  year={2025},
  volume={},
  number={},
  pages={1-6},
  doi={10.1109/WCNC.2007.207}}

@ARTICLE{Ishihara2021evolution,
  author={Ishihara, Takumi and Sugiura, Shinya and Hanzo, Lajos},
  journal={IEEE Access}, 
  title={The Evolution of Faster-Than-{Nyquist} Signaling}, 
  year={2021},
  volume={9},
  number={},
month={Jun.},
  pages={86535-86564},
  doi={10.1109/ACCESS.2021.3088997}}

@ARTICLE{xu2014spectrally,  
author={T. Xu and I. Darwazeh},
journal={IEEE Trans. Veh. Techn.},
title={Transmission Experiment of Bandwidth Compressed Carrier Aggregation in a Realistic Fading Channel},
year={2017},
volume={66},
number={5},
pages={4087-4097},
month={May},}

\end{document}